\documentstyle[aps,prl, preprint,epsfig]{revtex}

\begin{document}
\title{Magnetic--Field Induced Localization in the Normal State of Superconducting 
La$_{2-x}$Sr$_{x}$CuO$_4$}
\author{ A. Malinowski, $^1$ Marta Z. Cieplak,$^{1,2}$ A. S. van
Steenbergen, $^3$ J. A. A. J. Perenboom, $^3$ K. Karpi\'{n}ska,$^1$, 
M. Berkowski $^1$, S. Guha, $^2$ and P. Lindenfeld $^2$ }
\address{$^1$ Institute of Physics, Polish Academy of Sciences, 02 668 Warsaw,
Poland\\ 
$^2$ Department of Physics and Astronomy, Rutgers University, Piscataway, 
NJ 08855, USA\\
$^3$ Research Institute for Materials, High Field Magnet Laboratory,
University of Nijmegen, 6525 ED Nijmegen, 
The Netherlands}
\maketitle
\begin{abstract}

Magnetoresistance measurements of highly underdoped superconducting
La$_{2-x}$Sr$_x$CuO$_4$ films with $x = 0.051$ and $x = 0.048$, performed in
dc magnetic fields up to 20 T and at temperatures down to 40 mK, reveal
a magnetic--field induced transition from weak to strong localization
in the normal state. The normal--state conductances per CuO$_2$--plane,
measured at different fields in a single specimen, are found to
collapse to one curve with the use of a single scaling parameter that
is inversely proportional to the localization length. The scaling
parameter extrapolates to zero near zero field and possibly at a finite
field, suggesting that in the zero--field limit the electronic states
may be extended.

\end{abstract}
\pacs{74.72.Dn, 74.76.Bz, 74.20.Mn, 74.25.Fy }

\newpage

The unusual normal--state transport properties of high--$T_c$
superconductors include strongly anisotropic resistivities, a
temperature--dependent Hall effect, and anomalous magnetoresistance
\cite{iye}. The character of the electronic ground state underlying
superconductivity is the subject of experiment and speculation, 
and is expected to be different according to different
models, with suggestions that it is insulating \cite{ins}, or weakly localized
in two dimensions (2D) \cite{ande}, as recently reported in 
La$_{2-x}$Sr$_x$CuO$_4$ (LSCO) in very high magnetic fields \cite{ando}.

Extremely high magnetic fields are required to quench superconductivity
when the transition temperature is high. We have therefore investigated
strongly underdoped, but still superconducting specimens, with values
of $T_c$ reduced below 4 K, in magnetic fields up to 20 T, at
temperatures down to 40 mK. We show that the field localizes the
carriers, leading to variable--range hopping at the highest fields and
lowest temperatures, so that the behavior observed in strong fields is
not a reliable guide to the nature of the zero--field electronic
``ground state'' in the absence of superconductivity. 

We also show that the normal--state conductance per CuO$_2$ plane,
measured at different fields in one specimen, may be collapsed to a
single curve by adjusting a single scaling parameter. The scaling
indicates a gradual transition from weak localization at low fields 
to strong localization at high fields, similar to the
disorder--induced localization observed in conventional 2D and 3D
metals and semiconductors \cite{liu,hsu,mob,zhang}. 
However, within our experimental accuracy,
the scaling parameter $T_0$, which is inversely proportional to the
localization length, extrapolates to zero at fields that are close to
zero and possibly finite, suggesting that in the zero--field limit the
electronic state underlying superconductivity may be extended.

The specimens were $c$--axis aligned epitaxial films, grown by pulsed
laser deposition on SrLaAlO$_4$ substrates \cite{mar}. They were patterned by
photolithography, and silver pads were evaporated for four--point
resistivity measurements. The magnetoresistance (MR) measurements were
made in magnetic fields up to 20 T, generated by Bitter magnets, in two
different low--temperature setups to check for consistency. One was a
He$^3$ cryostat with dc measurements and temperatures down to 600 mK.
The other was a dilution refrigerator in which 3 Hz--ac was used, with
temperatures down to 40 mK. Most of the data were accumulated by
sweeping the field. Several runs were also made by sweeping the
temperature, and were found to be consistent with the others. The data
from the two setups differred by less than 5\%. This difference
reflects slightly different field calibrations and small differences
in current, and is insignificant for the discussion of this paper.

We measured two films with a nominal composition given by $x = 0.051$.
The values of $T_c$, $\rho$ and MR differred only slightly, so that we
present the results for only one of them. It is designated as specimen
S1, with a value of $T_c$ of 3.8 K, and $ab$--plane resistivity,
$\rho_{ab}$, at
40 K equal to 2.9 $m\Omega cm$. The temperature dependence of $\rho_{ab}$
in zero field is shown in the inset of Fig. 1. We also measured the MR
of a third film, S2, with nominal composition $x = 0.048$, and $T_c$ =
400 mK, which was measured earlier up to 6 T \cite{karp}.

In all cases the magnetic field was perpendicular to the $ab$--plane.
In Fig. 1 $\rho_{ab}$ for specimen S1 is plotted against $\ln{T}$ for
fields from 7 to 10.6 T, and in Fig. 2 for fields from 7 to 20 T.
It is apparent that the field gradually quenches superconductivity and
induces a superconductor--insulator transition, similarly to the behavior described
previously for specimen S2 \cite{karp}, except for the higher fields that are 
necessary to suppress superconductivity in S1. In Ref. \cite{karp} we analyzed
the nature of this S--I transition, and found that it differs from the
Cooper--pair localization predicted by Fisher \cite{fish}. 

The inset to Fig. 2 shows $\rho_{ab}$ for specimen S1, plotted as 
ln $\rho_{ab}$ against $T^{-1/4}$. For the highest fields the data follow
straight lines to $T^{-1/4}$ = 2 (corresponding to $T = 60$ mK), consistent
with 3D Mott variable--range hopping \cite{shklo}. The slopes of the straight lines
increase with increasing field, pointing to field--induced localization
of the carriers. In the field and temperature range of this experiment
the MR was positive for all films, approaching an approximately linear
dependence on field at the highest fields. This differs from the
results of Ref. \cite{ando} on single crystals of LSCO with $x = 0.08$
and 0.13, where the MR in the limit of high fields was found to be negative.

The saturation of $\rho_{ab}$ below 60 mK is presumably a result of
superconducting fluctuations. For lower fields the fluctuations occur
at higher temperatures, and the $T$--dependence of $\rho_{ab}$ becomes
weaker than exponential. It may be seen that for some fields and
temperatures the $T$--dependence is close to $ln(1/T)$, as observed in Ref.
\cite{ando} down to about 0.7 K. It is apparent, however, that this is only an 
intermediate
stage in the gradual evolution from variable--range hopping to weakly
localized and eventually metallic behavior, so that the logarithmic
dependence by itself does not seem to have any special significance.

Fig. 3a shows the data for film S1 as a log--log graph of the
conductance per single CuO$_2$ plane, $G$, against temperature, at
different fields. We now adopt the scaling procedure used in several 
previous studies of disorder--induced localization in various 2D and 3D
systems \cite{liu,hsu,mob,zhang}. 
We find that shifting the data for the different fields
along the $\ln{T}$--axis allows the collapse of the normal--state data to a
single curve, as shown in Fig. 3b. In this procedure we plot the data
against $\ln{{\alpha}T}$, where ${\alpha}(B)$ is set equal to one for 
$B = 20$ T, and
chosen for other fields so as to superimpose the curves, as on Fig. 3b.
The deviations on the low--$T$ side result from superconducting fluctuations.
The $\ln{G}$ scale is normalized by the constant $G_{00}$, equal to 
$e^2/2\pi^2\hbar$, to allow a direct comparison with the results of Ref.
\cite{hsu}, where $G_{00}$ was found to separate the strong and weak
localization regimes in metallic disordered 2D films. We find
variable--range hopping in the limit $G/G_{00} \ll 1$, changing to a weaker
$T$--dependence as $G/G_{00}$ approaches one. This behavior resembles the
transition from strong to weak localization described in 
Refs. \cite{liu} and \cite{hsu}.

The strong--localization region is characterized by the parameter
$T_0$ in the Mott variable--range hopping law, which is inversely
proportional to the localization length. Since the conductance depends
on temperature only in the combination $T_0/T$, a shift from $T$ to 
${\alpha}T$ 
is equivalent to a shift from $T_0$ to $T_0/\alpha$, so that 
$T_0(B) = T_0(20$T$)/\alpha$. We therefore use the factor ${\alpha}(B)$,
determined as the
factor in the scaling procedure that superposes the curves on Fig. 3b,
to determine also $T_0(B)$, even in the regime where the charge
carriers are no longer strongly localized and Mott's law no longer
applies. At some lower, possibly inaccessible temperature, Mott's law
can be expected to hold again, with this value of $T_0$.
This definition of $T_0$ allows us to plot the scaled conductance as a
function of $\ln{T/T_0}$, as on the upper scale of Fig. 3b. 

In Fig. 4 we plot the normal--state curve constructed in this way for
the two films S1 ($x = 0.051$) and S2 ($x = 0.048$). In order to cause the two
curves  to be superimposed to form a single curve, it is necessary
to rescale not only the horizontal, but also the vertical axis. This is
different from the case of disorder--induced localization in metallic 2D
films \cite{liu,hsu}, where the conductance approaches a constant value,
independent of disorder, in the high--$T$ limit. This difference may
be related to the fact that the metal--insulator transition in LSCO is
inherently different, apparently driven primarily by band filling
\cite{tokura}.

The figure also shows the data from Ref. \cite{ando}, for an LSCO crystal with
$x = 0.08$, in a pulsed field of 50 T, scaled to join the other curves at
the high--$T$ end of their data. It may be seen that for this
specimen even the 50 T--field does not induce strong localization, and
that the curve departs from the shape of the other two curves as the
temperature is lowered. If, as stated in Ref.\cite{ando}, the field is
sufficient to suppress superconducting fluctuations, the curve from
their data on Fig. 4 seems to indicate the likelihood of metallic
character for their specimen. This conclusion differs from that of Ref.
\cite{ando}, where specimens up to the optimally doped, i.e. for values of 
$x \leq 0.15$, are said to be insulating.  There are two reasons for the
difference in our conclusions. First, Ando {\it et al.} characterize a
specimen as ``insulating'' when the slope of $R(T)$ is negative, while we
use the much more stringent criterion that there must be a finite value
for $T_0$ and hence for the localization length.  Second, as we show,
the localization in the field does not necessarily imply localization in
the absence of a field.  These differences in interpretation are not
likely to be affected by differences in the specimen characteristics,
such as that implied by the negative MR of the specimen of Ref. \cite{ando},
which suggests differences in spin scattering, presumably resulting
from differences in structure and composition.

Strict adherence to scaling would lead to the conclusion that the
transition to strong localization must take place regardless of the
value of the magnetic field. It must be remembered, however, that the
scaling procedure superimposes curves for different fields, and is
valid only as long as the magnetic field dominates the localization.
The validity of the scaling procedure breaks down when the field
becomes so small that the magnetic length, $\sqrt{c\hbar/eH}$,
becomes larger than other scattering lengths, and other processes dominate.

In order to assess what happens at small fields we plot $T_0$ as a function
of $B$ for specimens S1 and S2 in Fig. 5, using the values determined by the
scaling procedure. The random errors associated with the scaling
procedure are about the size of the data points.  There is also a
systematic error resulting from the fit
of the hopping expression at the highest fields, indicated by the error
bars at the highest--field points. We find that the best description of
the field variation is given by the power law shown in Fig. 5 by the
dashed lines, $T_0(B) = A[(B - B_{cr})/B_{cr}]^\nu$, with the parameter
$B_{cr}$ equal to 2.5 $\pm$ 0.5 and 1.1 $\pm$ 0.2 for samples S1 and
S2 respectively, $A$ equal to 19.3 $\pm$ 3.9 and 65.7 $\pm$ 14.0, and 
$\nu$ essentially the same in both samples, 
0.64 $\pm$ 0.04  and 0.62 $\pm$ 0.06.
Taken literally this formula suggests that there is
a finite field, $B_{cr}$, below which $T_0$ is equal to zero, a field
below which the electronic states would be extended and at which a
metal--insulator transition would then take place. Experimentally the
region near $B_{cr}$ is masked by superconducting fluctuations, so that
the extrapolation must be treated with caution. Nevertheless this
analysis of the data suggests the possibility that in the zero--field
limit the electronic states in these two specimens may indeed be
extended.

In summary, we find that the magnetic field induces a transition from
weak to strong localization in LSCO, in many respects similar to the
disorder--induced transition in conventional disordered systems. The
extrapolation to zero magnetic field suggests that the extrapolated
electronic ``ground state'' in the absence of superconductivity is
extended rather than localized.

We would like to thank E. Abrahams, T. Dietl, and G. Kotliar for many
helpful discussions, and J. G. S. Lok for help with the MR
measurements.
This work was supported by the Polish Committee
for Scientific Research, KBN, under Grant 2P03B 05608, by the National
Science Foundation  under Grant DMR 95--01504, by the European Union
under Contract CIPA--CT93--0032, and by the FOM (The Netherlands) with
financial support from NWO.

\narrowtext

\begin{figure}[t]
\epsfig{file=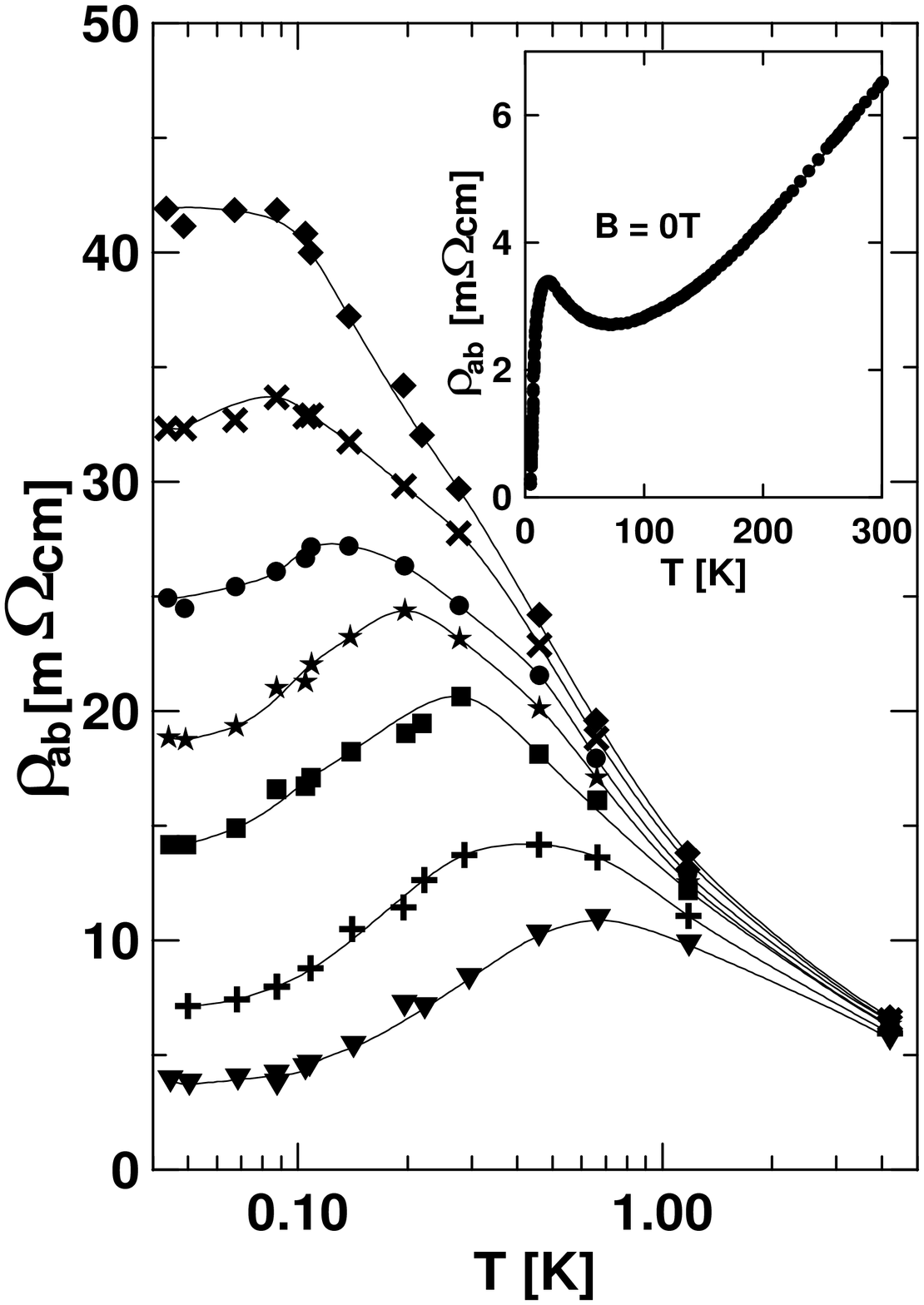,width=0.8\textwidth}
\caption{
The temperature dependence of $\rho_{ab}$ for film S1 in perpendicular 
magnetic fields. The fields are, from below, 7, 8, 9, 9.4, 9.8, 10.2,
10.6 T. The open circles between $T = 0.7$ K
and $T = 1$ K show the data obtained with dc as explained in the text.
The lines are guides to the eye.
The inset shows $\rho_{ab}(T)$ in zero field.}
\label{fig1} 
\end{figure}

\begin{figure}[t]
\epsfig{file=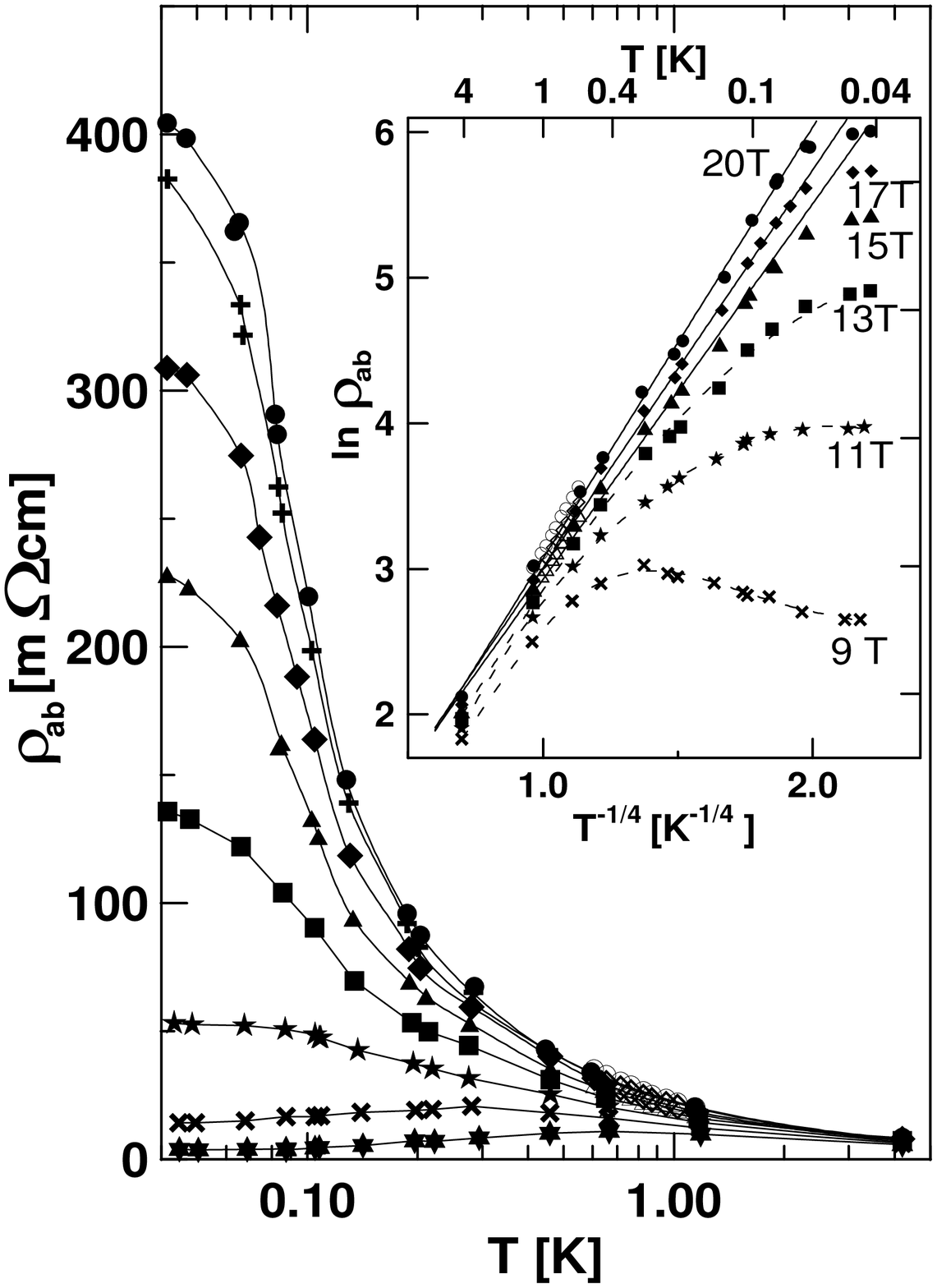,width=0.8\textwidth}
\caption{
The temperature dependence of $\rho_{ab}$ for film S1 in perpendicular 
magnetic fields. The fields are, from below, 7, 9, 11, 13, 15, 17, 19,
and 20 T. The open circles between $T = 0.7$ K
and $T = 1$ K show the data obtained with dc as explained in the text.
The lines are guides to the eye.
The inset shows $\ln{\rho_{ab}}$ versus $T^{-1/4}$ for several
magnetic fields.}
\label{fig2}
\end{figure}

\begin{figure}[t]
\epsfig{file=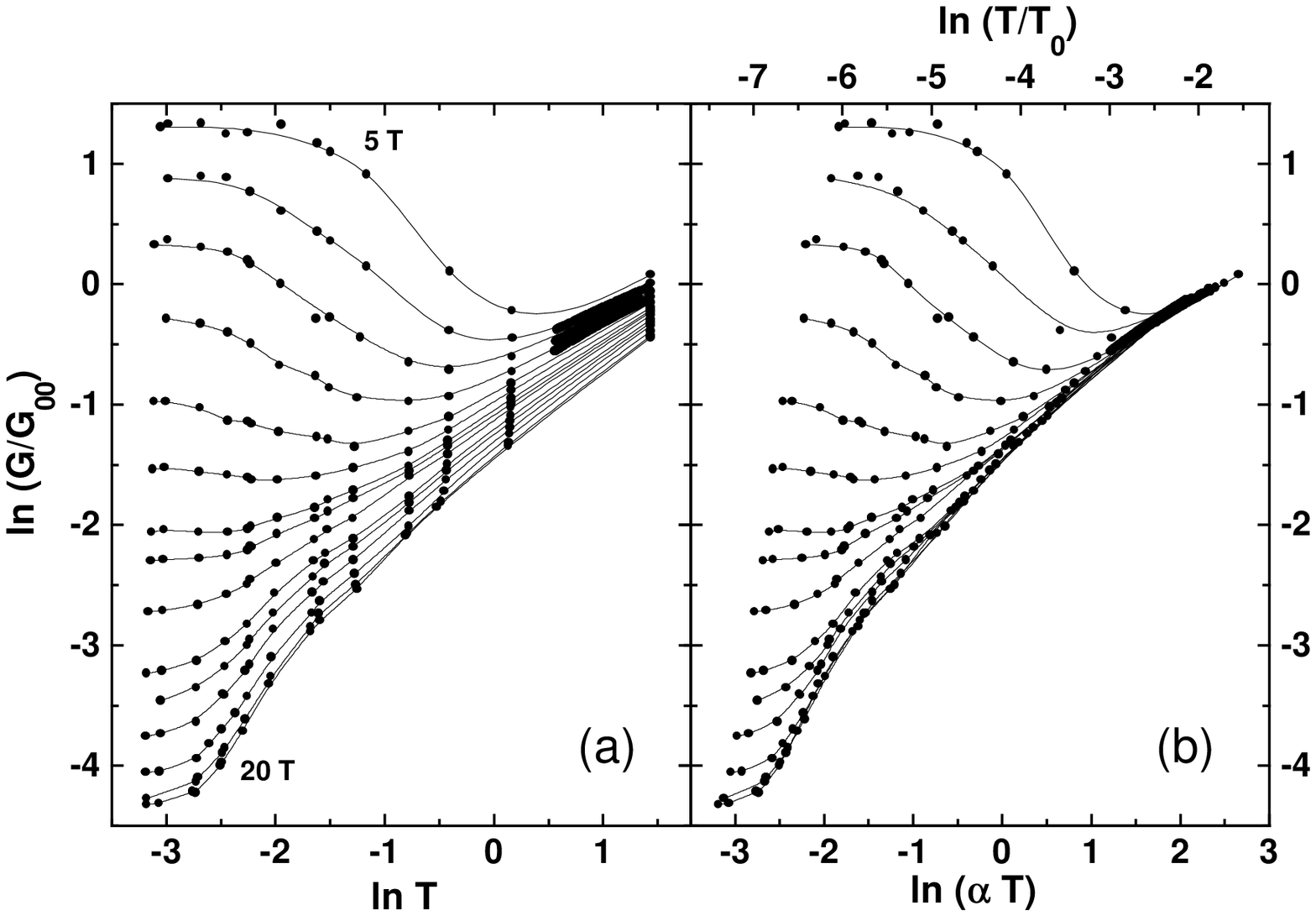,width=0.8\textwidth}
\caption{
Conductance per single CuO$_2$ plane, $G$, normalized to 
$G_{00} = e^2/2\pi^2\hbar$ for film S1 in magnetic fields ranging from 5 T
(topmost curve)  to 20 T (lowest curve), plotted (a) versus temperature
on a logarithmic scale, and (b) versus temperature scaled by the
factor ${\alpha}(B)$, or, equivalently (upper scale), by $T_0(B)$.}
\label{fig3}
\end{figure}

\begin{figure}[t]
\epsfig{file=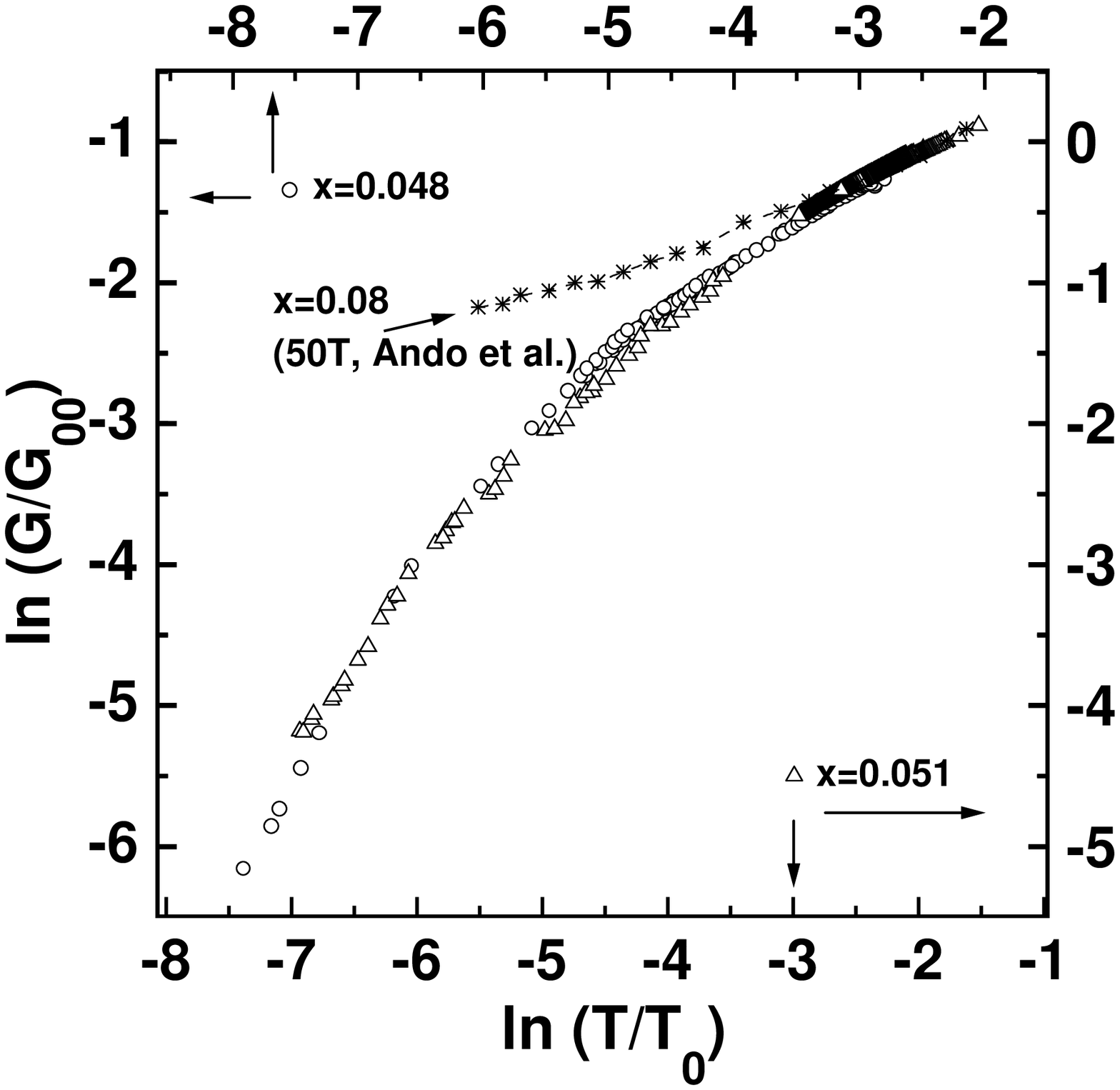,width=0.8\textwidth}
\caption{
Log-log graph of the scaled normal--state conductance as a
function of the scaled temperature for films
S1 and S2, and for the single crystal from Ref.\protect\cite{ando}.}
\label{fig4}
\end{figure}

\begin{figure}[t]
\epsfig{file=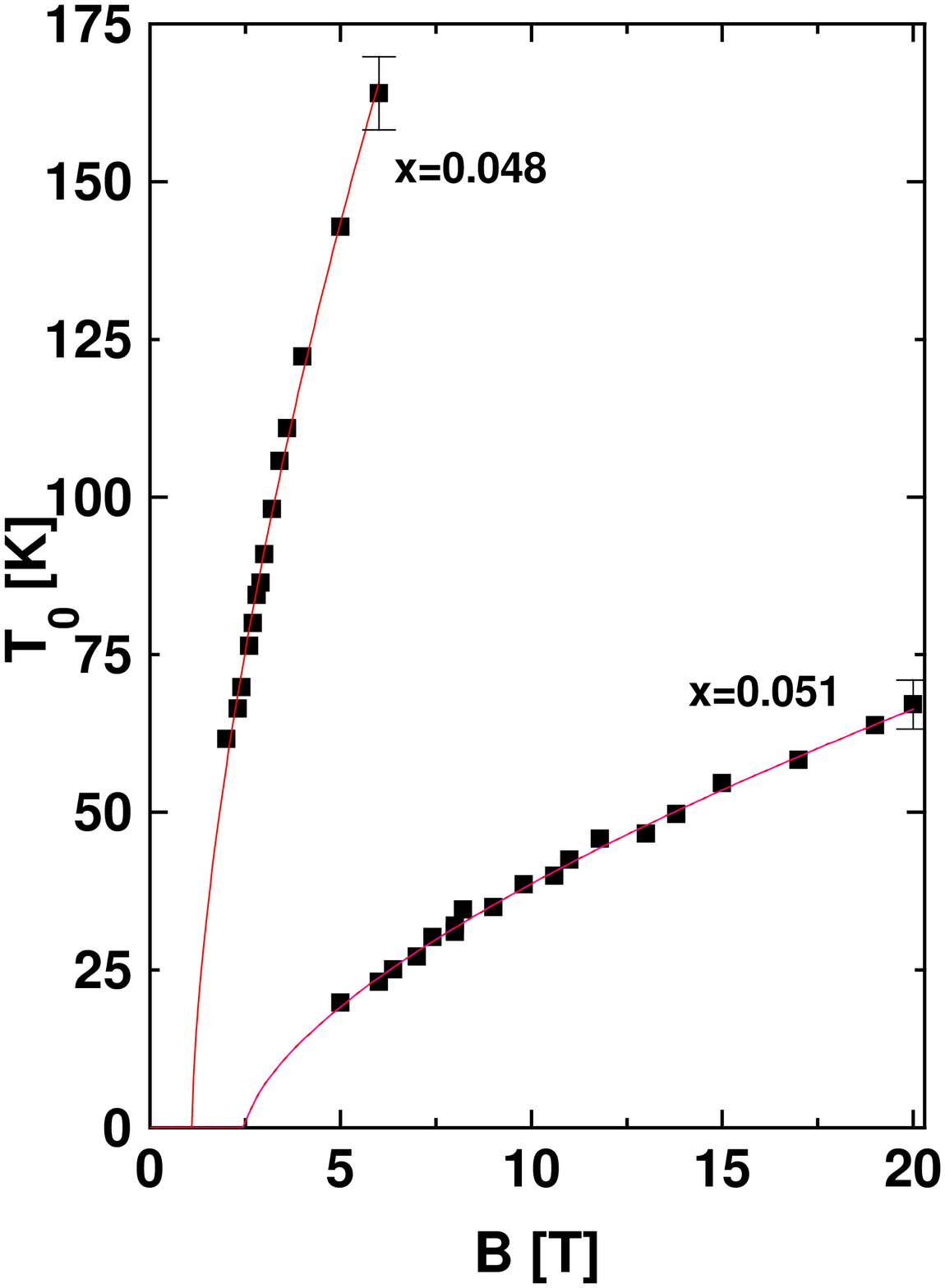,width=0.8\textwidth}
\caption{
The dependence of the parameter $T_0$ on magnetic field for
films S1 ($x=0.051$) and S2 ($x=0.048$). The error bars show the uncertainty
resulting from the fit of the Mott relation to the data for the highest
magnetic field. The dashed lines show the expression 
$T_0 = A [(B-B_{cr})/B_{cr}]^{\nu}$, fitted to the data, with the parameters
$A$, $B_{cr}$, and $\nu$, as described in the text.}
\label{fig5}
\end{figure}

\end{document}